\documentclass[aps, pra, reprint, twocolumn, flushbottom, superscriptaddress, floatfix]{revtex4-1}

\usepackage{graphicx}
\usepackage{gensymb}

\usepackage{placeins}

\date{\today}

\newcommand{\ort}{\mathbf{r}}

\newcommand{\enf}{\mathbf{E}_\mathrm{nf}}
\newcommand{\ein}{\mathbf{E}_\mathrm{in}}
\newcommand{\epsr}{\epsilon_\mathrm{r}}
\newcommand{\epsi}{\epsilon_\mathrm{i}}
\newcommand{\nm}{\mathrm{nm}}
\newcommand{\fs}{\mathrm{fs}}

\begin{document}

\title{Large optical field enhancement for nanotips with large opening angles}

\author{Sebastian Thomas}
\thanks{These authors contributed equally to this work.}
\affiliation{Department of Physics, Friedrich-Alexander-Universit\"at Erlangen-N\"urnberg, Staudtstra\ss e 1, D-91058 Erlangen, Germany, EU}
\author{Georg Wachter}
\thanks{These authors contributed equally to this work.}
\affiliation{Institute for Theoretical Physics, Vienna University of Technology, Wiedner Hauptstr. 8-10, A-1040 Vienna, Austria, EU}
\author{Christoph Lemell}
\affiliation{Institute for Theoretical Physics, Vienna University of Technology, Wiedner Hauptstr. 8-10, A-1040 Vienna, Austria, EU}
\author{Joachim Burgd\"orfer}
\affiliation{Institute for Theoretical Physics, Vienna University of Technology, Wiedner Hauptstr. 8-10, A-1040 Vienna, Austria, EU}
\affiliation{Institute of Nuclear Research of the Hungarian Academy of Sciences (ATOMKI), H-4001 Debrecen, Hungary, EU}
\author{Peter Hommelhoff}
\affiliation{Department of Physics, Friedrich-Alexander-Universit\"at Erlangen-N\"urnberg, Staudtstra\ss e 1, D-91058 Erlangen, Germany, EU}

\begin{abstract}
We theoretically investigate the dependence of the enhancement of optical near-fields at nanometric tips on the shape, size, and material of the tip. We confirm a strong dependence of the field enhancement factor on the radius of curvature. In addition, we find a surprisingly strong increase of field enhancement with increasing opening angle of the nanotips. For gold and tungsten nanotips in the experimentally relevant parameter range (radius of curvature $\geq 5\,\nm$ at $800\,\nm$ laser wavelength), we obtain field enhancement factors of up to ${\sim}35$ for Au and ${\sim}12$ for W for large opening angles. We confirm this strong dependence on the opening angle for many other materials featuring a wide variety in their dielectric response. For dielectrics, the opening angle dependence is traced back to the electrostatic force of the induced surface charge at the tip shank. For metals, the plasmonic response strongly increases the field enhancement and shifts the maximum field enhancement to smaller opening angles.
\vskip .5cm
{\footnotesize Email: sebastian.thomas@fau.de, georg.wachter@tuwien.ac.at}
\end{abstract}

\maketitle


\section{Introduction}

\begin{figure}[tb]
    \begin{center}
    \includegraphics[width=\columnwidth]{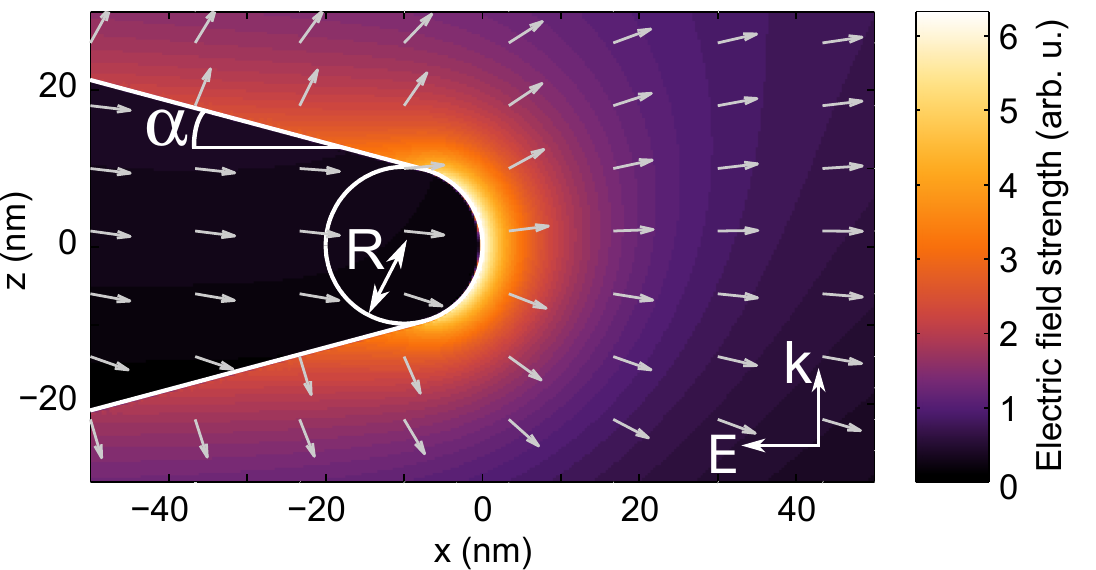}
    \end{center}
    \caption{Near-field of a $5\,\fs$, $\lambda = 800\,\nm$ laser pulse for an $R=10\,\nm$ tungsten tip with an opening angle of $\alpha=15\degree$. The laser pulse is propagating in the $z$ direction and is polarized along the $x$ direction. Shown here are the electric field strength (color) and the direction of the field (arrows) at the point in time when the near-field strength is at its maximum.
        }
    \label{tipfield}
\end{figure}

Optical near-fields arise when a structure illuminated by an electromagnetic wave is smaller than the wavelength of the impinging radiation. At the edges and protrusions of such a nanostructure, the electric field can be significantly enhanced. This nanoscale localisation of electric fields has recently found a large number of applications in nano-optics~\cite{Novotny2006a, Maier2007, Sarid2010}.
Due to the dynamic lightning rod effect that enables broadband field enhancement~\cite{Hartschuh2008, Martin2001, Goncharenko2006}, nano-sized tips are employed in a variety of applications such as scanning near-field optical microscopy (SNOM), tip-enhanced Raman scattering (TERS), and as sources of second-harmonic generation (SHG) or ultrafast photoemitted electrons~\cite{Novotny2006a, Hartschuh2008, Kawata2009, Bouhelier2003, Hommelhoff2006, Hommelhoff2006a, Ropers2007, Barwick2007}. The near-field enhancement and localisation at the apex of the nanotip play a key role in all these applications. Nonetheless, there is significant disagreement in the literature about the magnitude of the field enhancement at nanotips~\cite{Novotny2006a, Hartschuh2008}, most notably for gold tips where  theoretical and experimental results vary widely~\cite{Martin2001, Bouhelier2003, Ropers2007, Neacsu2005a, Arbouet2012, Thomas2013}.

Previous experimental and theoretical investigations have shown that details of the tip geometry near the apex can strongly influence the response~\cite{Martin2001, Neacsu2005a, Goncharenko2006a, Goncharenko2007, Behr2008, Zhang2009, Pors2014, Swanwick2014}. Even though modern nanofabrication techniques such as focused ion beam etching allow manufacturing of nanotips with custom-designed geometries, a systematic study of the relation between the tip design parameters (curvature, opening angle, and material) for realistic illumination conditions is still lacking. 

In this article, we investigate optical near-field enhancement at nano-sized tips as a paradigmatic example for a nano-structure. We perform fully three-dimensional (3d) numerical simulations employing Maxwell's equations combined with a realistic material-specific optical dielectric function $\epsilon(\omega)$ of nanotips as a pre-laboratory to guide optimization of the techniques that rely on localized field enhancement. We explore the dependence of optical near-field enhancement on the tip geometry for experimentally relevant tungsten and gold tips at $800\,\nm$ wavelength and a strong dependence on both the radius of curvature and the opening angle of the tip. We inquire into the origin of the unexpected field enhancement for larger angles for both materials. We generalize our results to a large class of materials by studying near-field enhancement as a function of the dielectric function of the tip material and find that increased field enhancement for larger angles persists for many materials and laser wavelengths. Technical details of the simulations as well as a comparison of nanotips to nano-ellipsoids, for which an analytical treatment is possible in the static limit, are given in the supplementary material. 

\section{Optical field enhancement at nanotips}
The contours of the near-field $|\enf(\ort)|$ follow the boundary of the nanostructure and the field strength decreases sharply with distance from the surface on the length scale of the radius of curvature $R$ of the nanostructure (see Fig.~\ref{tipfield}). For analytics and sensing applications, the most important property of near-fields is the strength of the enhanced near-field $|\enf|$ in comparison to the incident field $|\ein|$ described by the field enhancement factor $\xi$. Its magnitude can be quantified through
\begin{equation}
\xi = \max_{ \{ \mathbf r \} } \left\{ {|\enf(\ort)|}\,/\, {|\ein(\ort)|}  \right\} \, \, , 
\end{equation}
where the domain $\{\mathbf r \}$ extends over the entire region in the proximity of the nanostructure. Typically, the field enhancement is strongest on the surface of the nanostructure. 

Additionally, near-fields also feature a phase shift $\phi$ with respect to the exciting field. This can be expressed employing a generalized complex field enhancement factor $\xi = |\xi| \exp(i \phi)$~\cite{Bouhelier2003a}. When the field enhancement factor only weakly depends on the laser wavelength over the spectral width of the pulse, the phase shift $\phi$ is equivalent to  a shift of the carrier-envelope phase of few-cycle laser pulses. The latter becomes an important control parameter when the pulse duration is reduced to a few optical cycles as recently demonstrated in strong-field photoemission experiments from nanostructures~\cite{Kruger2011, Piglosiewicz2014}. 

\begin{figure*}[tb]
    \begin{center}
    \includegraphics[width=.8\textwidth]{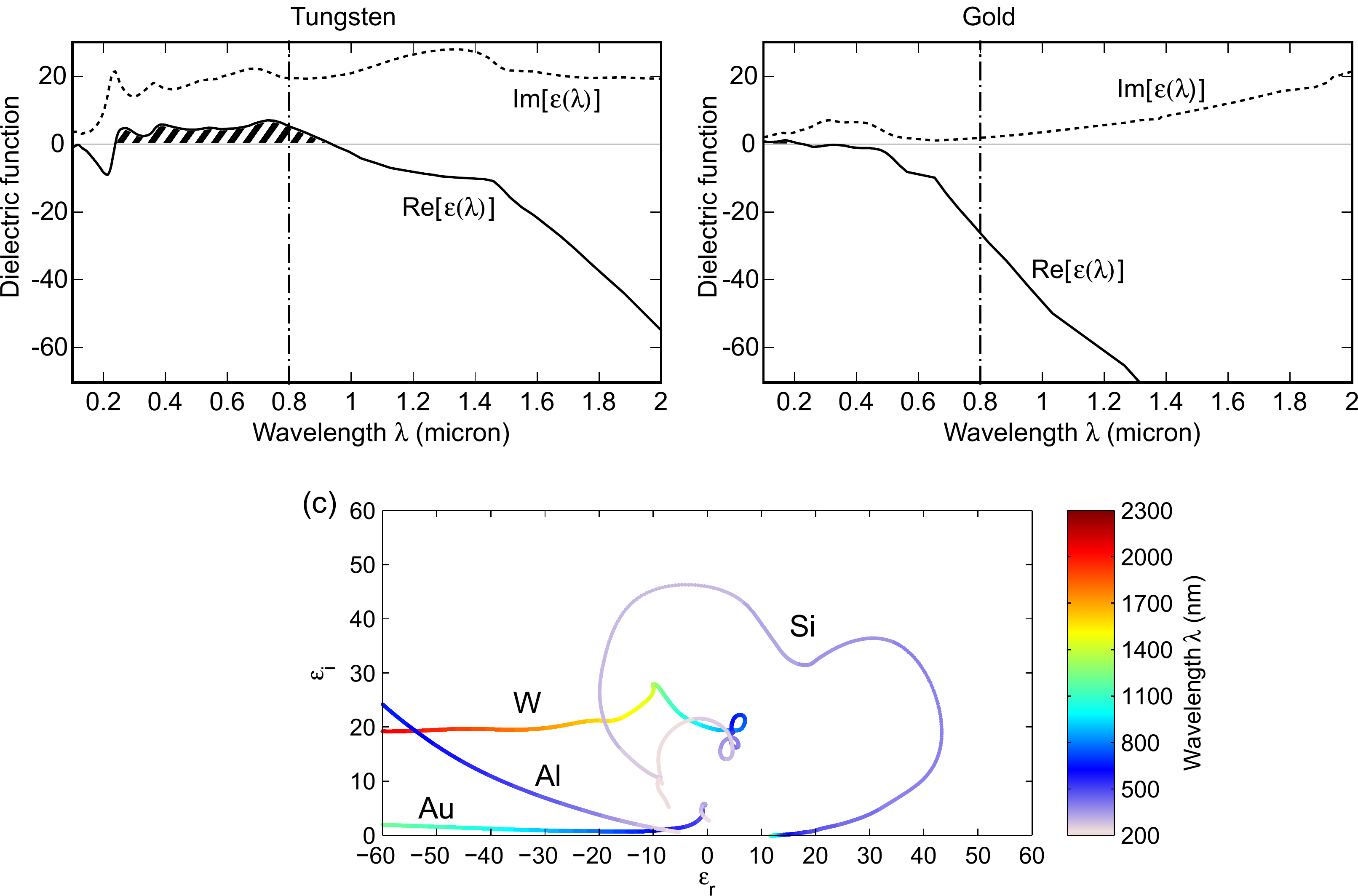}
    \end{center}
    \caption{
    Dielectric function of tungsten (a) and gold (b) between 100 nm and 2000 nm (vertical dash-dotted line: 800 nm). The real part of the dielectric function of gold is smaller than zero over most of the plotted range while tungsten has a positive dielectric function over a large wavelength range (hatched area). (c) shows the ``evolution'' of the complex dielectric function $\epsilon = \epsr + i\epsi$ of some typical nanotip materials in the $\epsr$-$\epsi$-plane with the wavelength as parameter (color box). Data for $\epsilon(\lambda)$ taken from Refs.~\cite{Lide2004, Palik1991}. 
    }
    \label{fig:dielfct}
\end{figure*}

To describe optical near-fields at nanotips, we consider a conical nanotip (Fig.~\ref{tipfield}) with a spherical cap at the apex located in the focus of a Gaussian laser beam. This corresponds closely to the geometry often used in photoemission and second-harmonic generation at nanotips. In SNOM and TERS experiments, the tip is typically close to a surface or another nanostructure, which can also contribute to, and usually increases, the field enhancement.

We numerically solve Maxwell's equations employing the finite-difference time-domain (FDTD) method, but cross-check our results with the boundary element method (BEM) as discussed in the supplementary material. The parameters that characterize our setup are:
\begin{itemize}
    \item the laser wavelength $\lambda$ and waist radius $w_0$ ($1/e^2$ intensity radius) of the focus,
    \item the radius of curvature $R$ and opening angle $\alpha$ of the tip (defined as the angle between the tip surface and its axis of symmetry, also called ``half-opening angle'', Fig.~\ref{tipfield}),
    \item and the optical properties of the tip material given by the frequency dependent dielectric function $\epsilon(\omega) = \epsilon_\mathrm{r}(\omega) + i \epsilon_\mathrm{i}(\omega)$ with $\epsilon_\mathrm{r(i)}$ the real (imaginary) part of $\epsilon(\omega(\lambda))$. 
\end{itemize}

\begin{figure*}[t]
    \begin{center}
    \includegraphics[width=.77\textwidth]{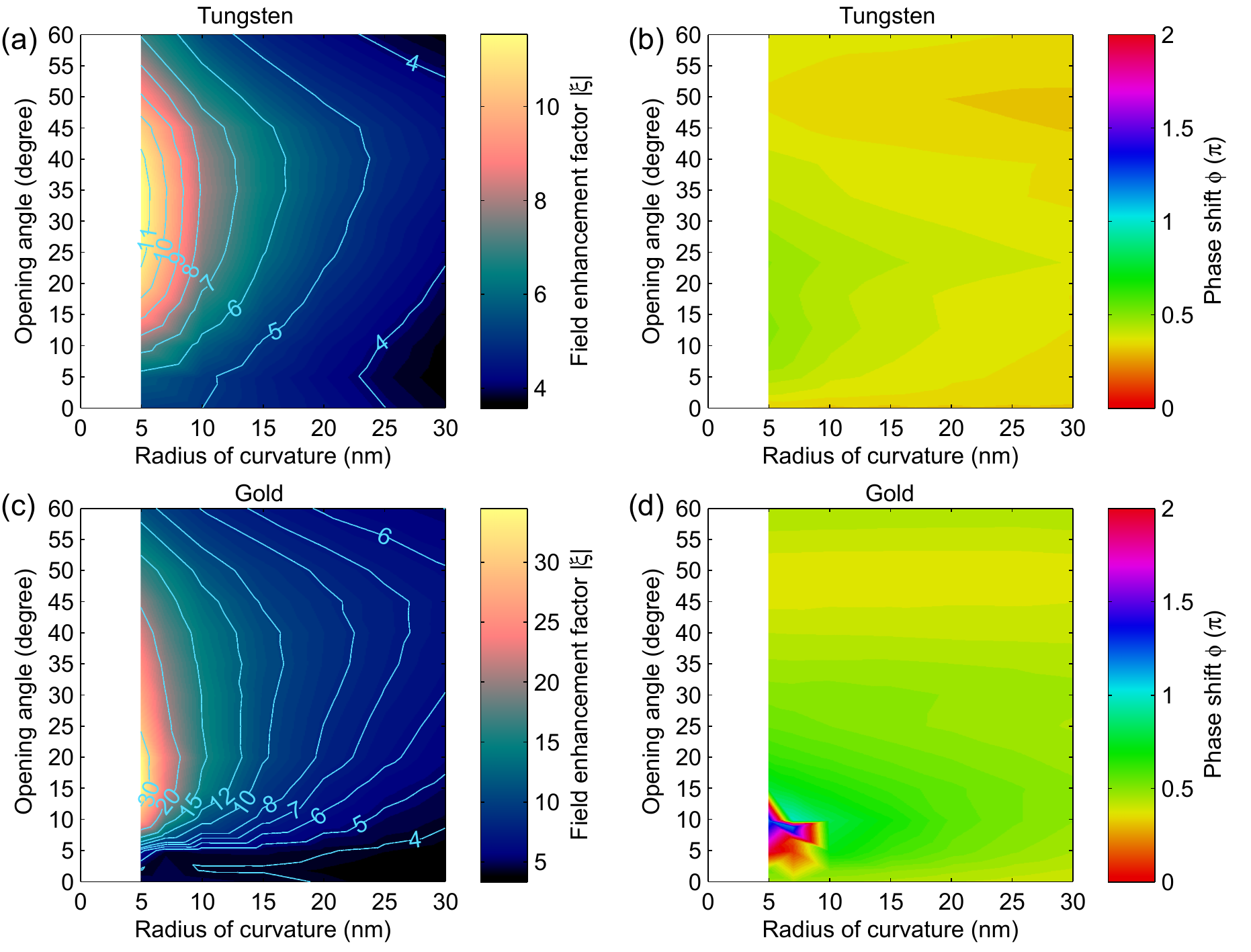}
    \end{center}
    \caption{Complex field enhancement factor $\xi = |\xi| \exp(i \phi)$ of tungsten (a,b) and gold tips (c,d) at $\lambda = 800\,\nm$ as a function of the radius of curvature of the tip and of the half-opening angle. Left column: $|\xi|$, right column: $\phi$. 
    }
    \label{angleradius}
\end{figure*}

As the laser beam waist is found not to significantly affect the field enhancement factor, the relevant parameters are reduced to $R$, $\alpha$, $\lambda$, and $\epsilon(\omega)$. Further, we may exploit the scaling invariance of Maxwell's equations~\cite{Joannopoulos2011}: an increase of the wavelength $\lambda \to \lambda' = s \lambda$ is equivalent to a decrease of the tip radius $R \to R' = R / s$ at the same value of the dielectric constant $\epsilon$. E.g., the field enhancement of a tip with $\tilde R = 20$ nm at a wavelength of $\tilde \lambda = 1600$ nm at dielectric constant $\tilde \epsilon = \epsilon(1600\,\nm)$ is the same as the field enhancement calculated for a tip of $R = 10$ nm at wavelength $\lambda = 800$ nm with the same dielectric constant $\tilde \epsilon$. We have numerically verified this scaling. In principle, this scaling property allows a further reduction of the parameter space. However, the required constancy of $\epsilon$ as a function of $\omega$ (or, equivalently, as a function of $\lambda = 2 \pi c / \omega$) imposes strong restrictions on realistic tip materials, and we hence do not exploit this scaling in the following simulations of gold and tungsten. Note that, while the maximum sharpness of the tip in applications is limited by the available fabrication technology, increasing the laser wavelength provides an attractive alternative to realize effectively sharper tips and thus obtain higher field enhancement.

In the following, we choose a fixed wavelength of $\lambda = 800\,\nm$ for which we have previously found good agreement between experiment and simulation for small opening angles $\alpha \lesssim 5 \degree$~\cite{Thomas2013} and discuss the effects of the remaining parameters $R$, $\alpha$ and $\epsilon$. One goal is to separate geometry effects from material effects. 

First, we investigate the influence of the tip geometry $(R, \alpha)$ on the field enhancement factor for two technologically relevant materials, tungsten and gold. At $\lambda = 800\,\nm$ wavelength, these materials show markedly different electromagnetic responses (Fig.~\ref{fig:dielfct}): The real part of the dielectric function is positive for tungsten ($\epsilon_\mathrm{W}(800\, \nm ) \approx 5+19i$) while it is negative for gold ($\epsilon_\mathrm{Au}(800\, \nm) \approx -23 + i$) \cite{Lide2004}. Tungsten thus behaves in the visible and near-infrared spectral region like a ``lossy'' dielectric with strong absorption as $\mathrm{Im}(\epsilon)$ is large. On the other hand, the negative dielectric function of gold, typical for metals, indicates plasmonic behavior. Corresponding eigenmodes, the surface plasmon polaritons (SPP), can be sustained at metal-dielectric interfaces. Their damping characterized by the small imaginary part of $\epsilon$ is weak compared to other nanotip materials.

The calculated field enhancement depends strongly on both the radius and the opening angle of the nanotip (Fig.~\ref{angleradius}). For both materials, the maximum enhancement is observed for small radii of curvature as expected for the dynamic lightning rod effect that predicts a field enhancement near sharp geometric features. Somewhat unexpectedly, however, we also find a strong dependence of the field enhancement on the tip opening angle for both materials. While the two materials display a similar field enhancement for small opening angles ($\alpha \leq 5\degree$) in agreement with recent experiments \cite{Thomas2013}, at intermediate opening angles ($10 \degree \lesssim \alpha \lesssim 40 \degree$) the field enhancement is further enhanced. This enhancement is more pronounced for gold tips than for tungsten tips. Gold tips display a distinct maximum enhancement at $\alpha \approx 15 \degree$. For tungsten, the maximum of the field enhancement is much broader and located around $\alpha \approx 40 \degree$. For $R = 5\,\nm$, the field enhancement factor can reach $|\xi|=36$ for gold tips near $\alpha=15\degree$ and $|\xi|=12$ for tungsten tips with $\alpha=35\degree$. For a larger radius of $R = 30\,\mathrm{nm}$, the dependence on the opening angle is weaker but still substantial with the maximum located near $\alpha \approx 45 \degree$ for both materials.

The phase shift also depends on both the opening angle and tip radius and is larger for gold tips than for tungsten tips. We observe the largest phase shift at intermediate angles $10 \degree \leq \alpha \leq 30 \degree$ for both materials. 
We find the absolute value of the field enhancement factor to be robust under variation of the details of the simulation while the phase shift is more sensitive (see the supplementary material for details). In the region where the strongest increase of field enhancement is observed for very sharp tips, we were not able in all cases to reliably extract the phase shift from the gold simulations (for $0 < \alpha \le 10 \degree$ and $R \le 10\, \nm$, Fig.~\ref{angleradius}(d)). We presume that this is due to a localized surface plasmon mode at the tip apex (see below). 

In order to explore the generality of the observed enhancement at large opening angles we varied the underlying tip geometry and considered paraboloid and hyperboloid tips. 
Paraboloid tips are defined entirely by the radius of curvature with their surface given by $x(y,z) = -(y^2+z^2)/(2R)$. For gold and tungsten paraboloids with $R = 5\,\nm$ to $30\,\nm$, the field enhancement is similar to conical tips for the same radius of curvature and opening angles around ${\sim}10\degree$. For hyperbolic tips, on the other hand, the radius of curvature and the asymptotic opening angle are independent parameters. There, we find that the field enhancement factor for a given radius of curvature depends significantly less on the opening angle than for conical tips. For $R=10\,\nm$ gold hyperboloids we obtain a field enhancement factor of ${\sim}10$ independent of the opening angle. This is because, for a constant radius of curvature, the asymptotic opening angle of a hyperbolic tip has only a weak effect on the shape close to the apex and only determines the shape of the shaft far away from the apex. This indicates that the field enhancement factor depends crucially on the tip shape in the vicinity of the apex, which provides clues as to its origin. 

\section{Model for the opening angle dependence of the field enhancement}

\begin{figure*}[bt]
    \begin{center}
    \includegraphics[width=.77\textwidth]{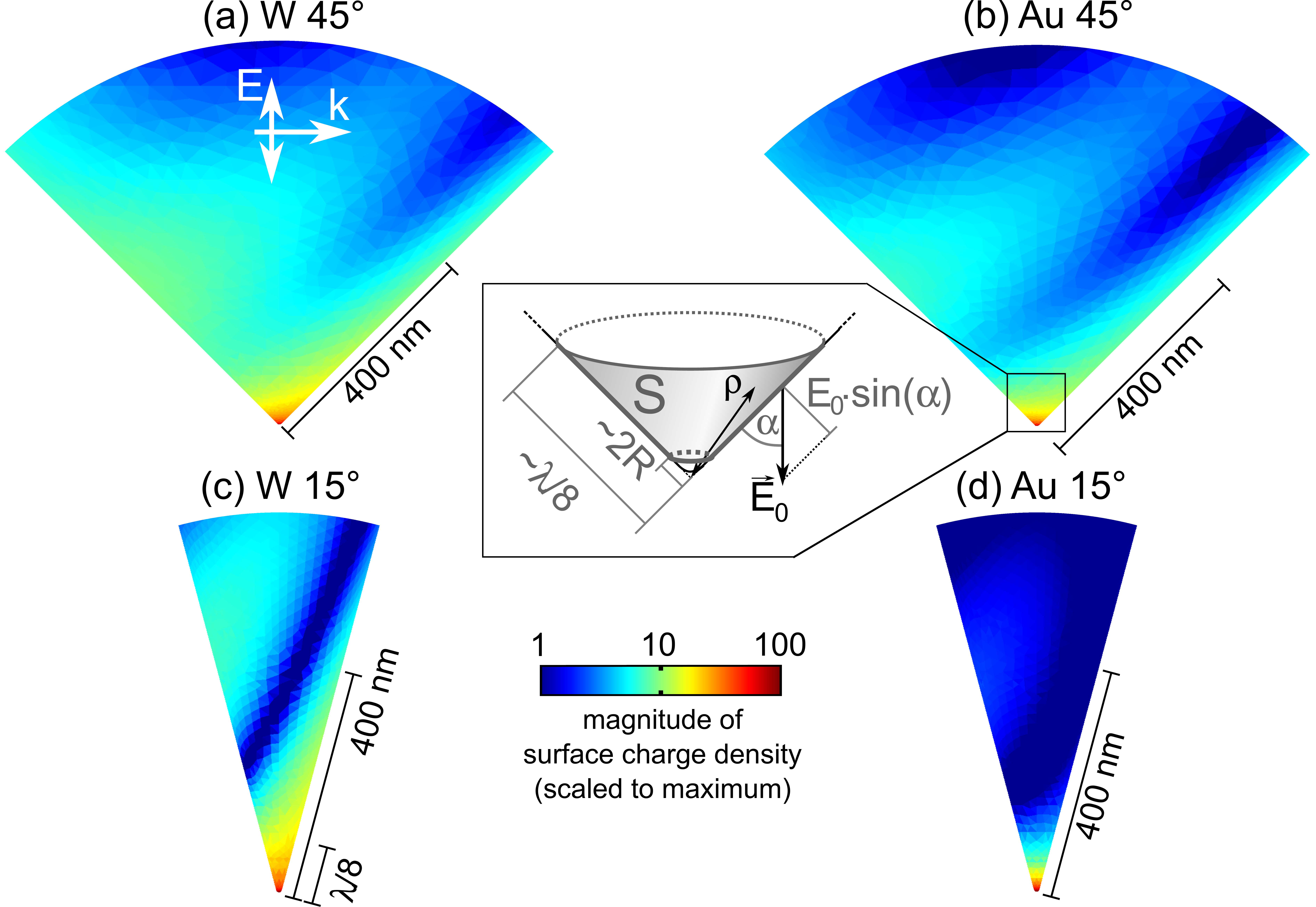}
    \end{center}
    \caption{
      Absolute magnitude of the surface charge density distribution on the nanotip near the apex calculated with the boundary element method. Laser propagation direction from left to right and polarization along tip axis. All tips have a tip radius of $R = 5 \, \nm$. Side view. (a) tungsten tip, $\alpha = 45 \degree$; (b) gold tip, $\alpha = 45 \degree$; (c) tungsten tip, $\alpha = 15 \degree$; (d) gold tip, $\alpha = 15 \degree$.  Inset (b): Coordinates for electrostatic model (Eq.~\ref{eq:elstat}). 
     }
    \label{fig:surfcharge}
\end{figure*}

We turn now to the modeling of the surprising increase of field enhancement with increasing opening angles. 
The first key observation is that the main contribution to the field enhancement at the apex is due to the electrostatic force exerted by the surface charge distribution in a small region around the tip apex (see Fig.~\ref{fig:surfcharge}) for all tip radii, opening angles, and tip materials, indicating that retardation effects on the micrometer length scale play only a minor role. This is in agreement with the work of van Bladel~\cite{VanBladel1996} and Goncharenko et al.~\cite{Goncharenko2006}.   

\begin{figure*}[tb]
    \begin{center}
    \includegraphics[width=.77\textwidth]{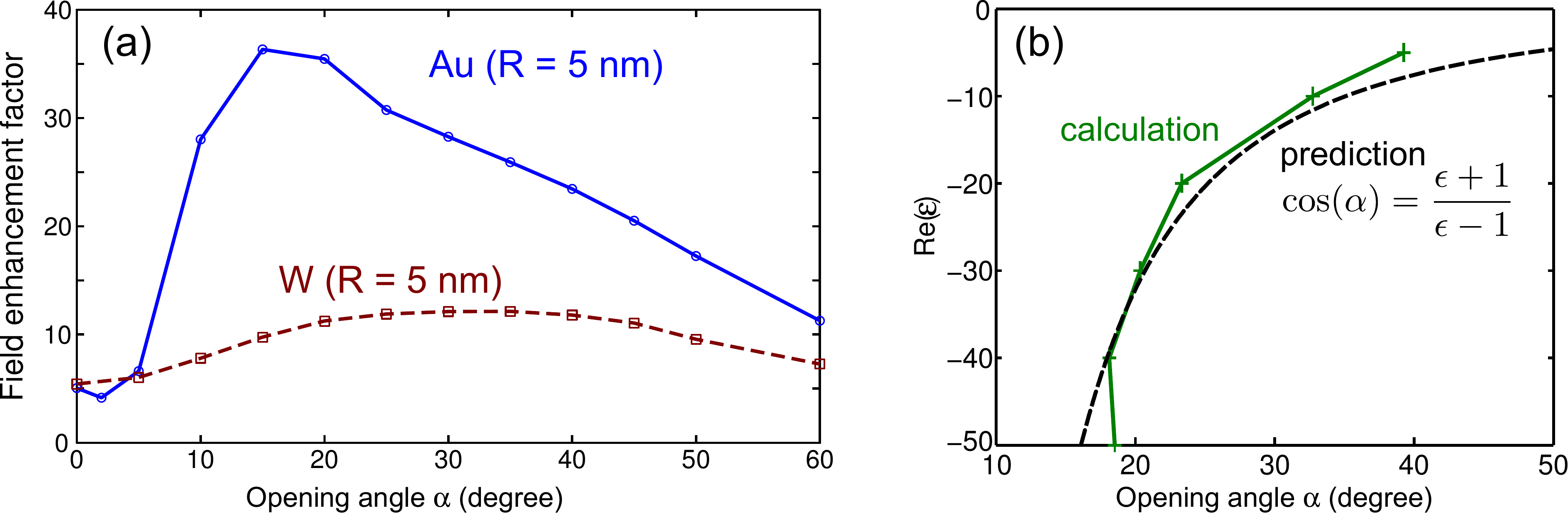}
    \end{center}
    \caption{
(a) Field enhancement factor as a function of tip opening angle for gold (blue solid line, circles) and tungsten tips (dark red dashed line, squares) with tip radius $R=5\,\nm$. (b) Maximum field enhancement factor as function of the real part of the dielectric function from FDTD simulations ($\mathrm{Im}(\epsilon)=5$, $R=10\,\nm$, green solid line and crosses), resonance angle according to Eq.~\ref{eq:sppres} (dashed line). 
}\label{fig:angle}
\end{figure*}

Focusing on the mechanism of field enhancement for tungsten and other dielectric materials ($\mathrm{Re}(\epsilon) > 0$), we find that the charge density distribution along the tip shaft is similar for all opening angles (e.g.~Fig.~\ref{fig:surfcharge}a,c), extending about $100\, \nm \approx \lambda/8$ along the tip shaft. The effect of this induced surface charge along the tip shank on the enhanced near-field at the apex may be investigated within an electrostatic model. Assuming for simplicity the magnitude of the induced surface charge to be constant along the tip shank near the apex in a region of size ${\sim}\lambda/8$ and proportional to the electric field strength perpendicular to the tip surface, the tip angle dependence of the surface charge is $\sigma_0(\alpha) \propto \sin(\alpha)$ (Fig. \ref{fig:surfcharge} inset). The contribution of the tip shank towards the field enhancement at the apex is 
\begin{equation}
\label{eq:elstat}
E^{\mathrm{apex}}(\alpha) \approx \int_S d^2 S \,\,\, \sigma_0(\alpha) \,\, \frac{1}{\rho^2}  \quad .
\end{equation}
The integral is taken over the surface $S$ of the tip shank from a lower limit near the tip apex ($\rho \gtrsim 2 R$) to an upper limit a fraction of the wavelength away from the apex ($\rho \lesssim \lambda/8$), where $\rho$ is the distance from the apex to a point on the tip surface (see Fig.~\ref{fig:surfcharge} inset). $E^\mathrm{apex}(\alpha)$ increases with increasing opening angle because the incident field component perpendicular to the tip surface increases. 
Eq.~\ref{eq:elstat} yields an angle-dependent component of the field enhancement 
\begin{equation} \label{eq:eapex}
  E^\mathrm{apex}(\alpha) \propto \sin^2(\alpha) \cos^2(\alpha)  \, \propto \sin^2(2 \alpha) \,\, . 
\end{equation}
While the details of the angular variance depend on the assumptions for the surface charge distribution and the shape of the surface $S$,  Eq.~\ref{eq:eapex} qualitatively describes the observed dependence for dielectrics. 
This model predicts a slow rise to a maximum field enhancement around $45 \degree$ in good qualitative agreement to the full calculations for tungsten where we find the maximum around $35 \degree$--$40 \degree$ (Fig.~\ref{fig:angle}a). We thus interpret the field enhancement for dielectrics as a geometrical effect that relies on the interplay between magnitude of induced surface charge $\sigma_0(\alpha)$ and the distance of the induced surface charge from the apex. 

\begin{figure*}[p]
    \begin{center}
    \includegraphics[width=0.8\textwidth]{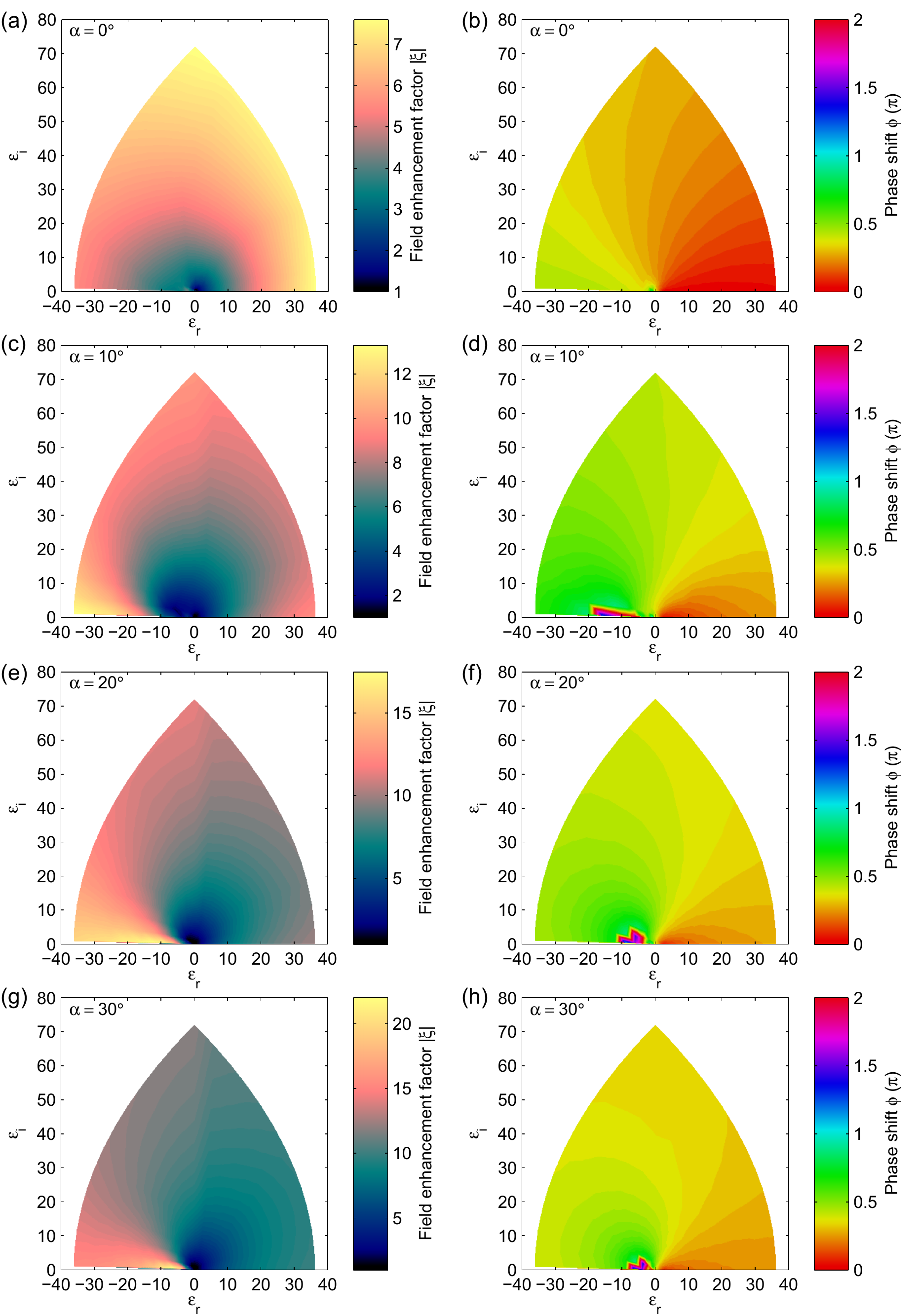}
    \end{center}
    \caption{Complex field enhancement factor $\xi = |\xi|\exp(i \phi)$ of $R = 10\,\mathrm{nm}$ tips at $\lambda = 800\,\mathrm{nm}$ as a function of the tip's dielectric constant for opening angles $0\degree$ (a,b), $10\degree$ (c,d), $20\degree$ (e,f), and $30\degree$ (g,h) for a selected region in the $\mathrm{Re}(\epsilon)$, $\mathrm{Im}(\epsilon)$ plane covering the range of dielectric functions of many materials at optical wavelengths (see Fig.~\ref{fig:dielfct}~(c)). Left column: $|\xi|$, right column: $\phi$.}
    \label{epsxi}
\end{figure*}

For plasmonic materials such as gold with $\mathrm{Re}(\epsilon)<0$, the induced surface charge at large tip opening angles resembles the result for dielectric tips (Fig.~\ref{fig:surfcharge}b), indicating a qualitatively similar mechanism of field enhancement at large angles. However, the maximum field enhancement is attained at a smaller opening angle, and the maximum is narrower than for dielectric materials (Fig.~\ref{fig:angle}a), pointing to an additional enhancement contribution at small angles \emph{and} small tip radii that is not present for dielectrics. 
At tip angles near the maximum field enhancement, our simulations show that the charge density distribution along the tip shaft is strongly localized at the apex (Fig.~\ref{fig:surfcharge}d), dominating the more extended pattern of the surface charge found for tungsten tips and larger angles.  
This suggests that the incident field couples to a surface plasmon mode localized at the tip apex causing the strong enhancement. 
The importance of surface plasmons for the observed dependence of field enhancement on the tip angle is corroborated by earlier work on near-field enhancement at the apex of a nanotip~\cite{Issa2007} as a result of adiabatic nano-focusing of surface plasmons along the shaft~\cite{Babadjanyan2000, Stockman2004}. 
While these observations pertain to a scenario with no external field present, their similarity to the present case of the amplification of an external field suggests that surface plasmons may also play a crucial role for the field enhancement. 

For a flat interface between a Drude metal with plasmon frequency $\omega_p$ (dielectric function $\epsilon_\mathrm{Drude}(\omega) = 1 - \omega_p^2 / \omega^2$) and vacuum ($\epsilon_\mathrm{vac} = 1$), the resonance condition for the well-known Ritchie surface plasmon~\cite{Ritchie1957} at frequency $\omega = \omega_p / \sqrt{2}$ is given by
\begin{equation} \label{eq:ressppdrude}
   \epsilon_\mathrm{Drude}(\omega) = 1 - \frac{ \omega_p^2 }{ \omega^2 } = -1 \,\,\, .
\end{equation}
The generalization of Eq.~\ref{eq:ressppdrude} to a cone with semiangle $\alpha$, infinitely sharp tip ($R \to 0$), and dielectric function $\epsilon(\omega)$ reads~\cite{Goncharenko2006, Vincent2009, Vincent2011}
\begin{equation} \label{eq:sppreseps}
  \epsilon(\omega) = \frac{ \cos( \alpha ) + 1 }{ \cos( \alpha ) - 1 }  \,\, .
\end{equation}
Eq.~\ref{eq:sppreseps} provides the link between the resonance frequency $\omega$, the frequency-dependent dielectric function $\epsilon(\omega)$ of the material, and the geometry of the tip described by the opening angle $\alpha$. Eq.~\ref{eq:sppreseps} can be equivalently written as
\begin{equation}
\label{eq:sppres}
\cos( \alpha ) = \frac{ \epsilon(\omega) + 1 }{ \epsilon(\omega) - 1 } \, \, .
\end{equation}
This resonance condition cannot be satisfied for dielectric tips where $\mathrm{Re} (\epsilon) > 0$ for any tip geometry as the right hand side is $> 1$. However, for gold at 800 nm, $\mathrm{Re}(\epsilon) = -23$ and the right-hand side of Eq.~\ref{eq:sppres} predicts a resonance around $\alpha = 23 \degree$ in good agreement to our simulations (Fig.~\ref{fig:angle}a). For materials in the infrared where $\mathrm{Re} (\epsilon) \to -\infty$ (compare Fig.~\ref{fig:dielfct}), the optimal angle approaches $0 \degree$. We confirm that the localized surface plasmon predicted by Eq.~\ref{eq:sppres} is indeed responsible for the field enhancement in our simulations by comparing the resonant angle $\alpha(\epsilon)$ predicted by Eq.~\ref{eq:sppres} with the angle for the maximum field enhancement found in our simulations as a function of the real part of the dielectric function (Fig.~\ref{fig:angle}b). We find overall good agreement between Eq.~\ref{eq:sppres} and our simulations whenever $\mathrm{Re} (\epsilon) < 0$. The results of our simulations are nearly independent of the precise value of $\mathrm{Im}(\epsilon)$ provided it is small, $\mathrm{Im}(\epsilon) / |\epsilon| \ll 1$.

A simple and transparent picture of field enhancement at nanotips thus emerges: For nanotips with large opening angles, the induced surface charge along the tip shank gives rise to a maximum around $\alpha = 45 \degree$ that can be understood from electrostatics. For plasmonic tips with $\mathrm{Re}(\epsilon) < 0$, an additional contribution arises from a localized surface plasmon mode at the tip apex, leading to even higher field enhancement and a sharper maximum at smaller angles.

\section{The dependence on the dielectric function}
To extend our results from tungsten and gold to other materials, we performed simulations varying the real and imaginary parts of the dielectric function of the tip material (Fig.~\ref{epsxi}). We fixed the tip radius at $R=10\,\mathrm{nm}$ and varied the opening angles between $0\degree$ and $30\degree$. The field enhancement factor increases with increasing tip opening angle for any given value of the dielectric function. However, as a function of $\epsilon$, $\xi(\epsilon)$ varies significantly for a given opening angle. For slim tips ($\alpha = 0\degree$, Fig.~\ref{epsxi}a), the field enhancement increases with increasing absolute value of the dielectric constant $|\epsilon|$. For $\alpha \ge 10\degree$, the field enhancement has a sharp maximum at negative real values of the dielectric function, for example at $\epsilon \approx -10 + 0 i $ for $\alpha = 30 \degree$ (Fig.~\ref{epsxi}g). This is interpreted in terms of the plasmon resonance expected around $\mathrm{Re}(\epsilon) = -14$ for $\alpha = 30 \degree$ (Eq.~\ref{eq:sppreseps}). With decreasing tip angle $\alpha \to 0$, Eq.~\ref{eq:sppreseps} predicts that this resonance moves towards $\mathrm{Re}(\epsilon) \to -\infty$, and we qualitatively observe that the maximum field enhancement and phase shift moves along the $\mathrm{Im}(\epsilon) = 0 $ axis towards $\mathrm{Re}(\epsilon) \to -\infty$ with decreasing tip opening angle. Therefore, and at first glance surprisingly, the plasmon resonance does not play a significant role for tips with very small opening angles below $5 \degree$ and for small absolute values of the dielectric function $|\epsilon|$ found for materials in the optical wavelength range (Fig.~\ref{fig:dielfct}c). This is the reason why the enhancement factors for plasmonic and dielectric materials closely resemble each other for small opening angles.

The results from Figs.~\ref{angleradius} and~\ref{epsxi} can be used to roughly estimate the field enhancement factor for other tip materials, radii $R'$ and wavelengths $\lambda'$ than those discussed here. First, one needs to obtain $\epsilon$ for the material and wavelength in question and look up the resulting $\xi$ from Fig.~\ref{epsxi} for the right opening angle. The so obtained result, however, is only correct for an effective tip sharpness $\kappa = \lambda/R = 800\,\nm/10\,\nm$. The behavior of $\xi$ for a different sharpness $\kappa'=\lambda'/R'$ can be approximated by scaling $\xi$ based on Fig.~\ref{angleradius}, where the field enhancement factor at $R=10\,\nm$ should be compared to an effective radius of $800\,\nm/\kappa'$. Depending on how far $\epsilon$ and $\kappa'$ are from the parameters discussed in this article, the resulting $\xi$ can be a good approximation or it may only indicate a trend.

\section{Conclusion}
We have explored the material and geometry dependence of optical near-field enhancement at nanostructures with the nanotip geometry taken as the prototypical example. We have discovered that, somewhat counterintuitively, larger field enhancement can be achieved for larger half-opening angles ($20 \degree$ to $40 \degree$) of the tip. This enhancement for fixed radius of curvature was found for both tungsten, exemplifying a dielectric response, and gold, a plasmonic material. Two processes contributing to this enhancement could be identified: 
For large opening angles, the increase of field enhancement can be understood from the electrostatic force of the induced surface charge along the tip shank.
This mechanism is effective in both dielectric and plasmonic materials. 
For the latter, excitation of localized surface plasmons at the apex gives rise to even stronger enhancement at intermediate angles. 
Varying the real and imaginary part of the dielectric function, we found the same qualitative behavior for a large number of materials, including other practically relevant materials such as aluminum, iridium, palladium, platinum, silicon, and silver. Our results indicate that, compared to currently employed tip shapes, a further field enhancement of magnitude 2 to 4 is achievable by employing tips with larger opening angles. We expect that such tips will provide a substantially increased signal especially for non-linear applications. 

The strong dependence of the enhancement on the tip geometry and not just on the radius of curvature may explain the many different values for the field enhancement factor of gold tips that have been reported in the literature, especially considering that the realistic shape of nanotips is more irregular than the conical tips employed in our simulations. 
The increase of field enhancement up to an optimal angle of $20 \degree$--$40 \degree$ depending on the tip material has escaped earlier studies \cite{Martin2001} presumably because the dependence on the opening angle was not sampled in sufficiently fine resolution. Our results suggest that higher field enhancement factors $|\xi|>10$ should be possible even for tungsten tips and other dielectric materials. This is consistent with a recent report of a field enhancement factor of ${\sim}10$ for silicon tips with a large opening angle~\cite{Swanwick2014}.
One reason why we did not observe higher field enhancements in our previous experiments with tungsten \cite{Thomas2013, Schenk2010, Wachter2012} may be related to the etching method we use for tungsten tips, which results in a small opening angle~\cite{Klein1997}. 

Our results may have ramifications for scanning near-field optical microscopy, tip-enhanced Raman spectroscopy and other techniques that rely on large field enhancement factors at rugged tips. Modern nanofabrication techniques such as focused ion beam etching could easily lead to the desired tip shape and larger enhancement factors.

\section*{Funding information}
This work was supported by the Gordon and Betty Moore Foundation, the ERC Grant ``NearFieldAtto'', FWF (Austria), SFB-041 ViCoM, SFB-049 Next Lite, and P21141-N16. G.W. thanks the International Max Planck Research School of Advanced Photon Science for financial support.

\section*{Acknowledgements}
We thank M.~T.~Homer Reid for fruitful discussions. Part of the calculations were performed using the Vienna Scientific Cluster (VSC).

\providecommand{\newblock}{}

\end{document}


\title{Large optical field enhancement for nanotips with large opening angles: supplementary material}

\maketitle

\manuallabel{main-tipfield}{1}
\manuallabel{main-eq:sppres}{6}
\manuallabel{main-eq:sppreseps}{5}
\manuallabel{main-epsxi}{6}
\manuallabel{main-fig:angle}{5}
\manuallabel{main-fig:dielfct}{2}

\section{FDTD: simulation setup}
\label{fdtdsetup}

\begin{figure}[b]
    \begin{center}
    \includegraphics[width=.9\columnwidth]{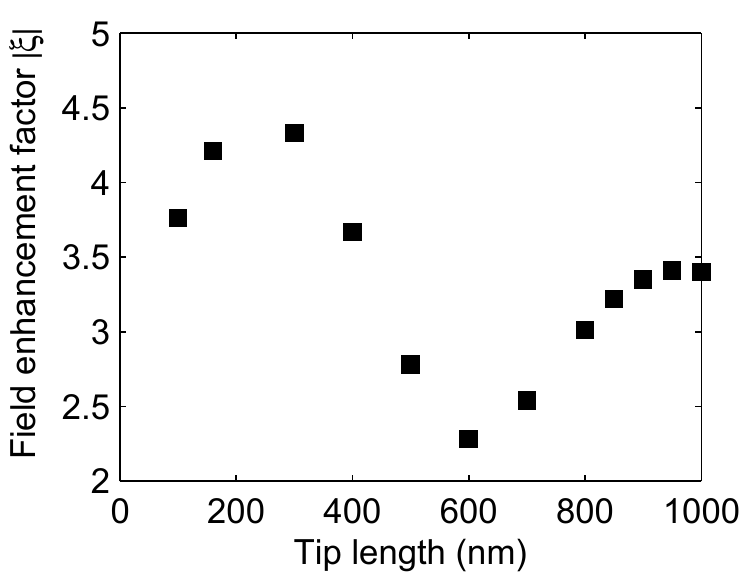}
    \end{center}
    \caption{Extracted field enhancement factor as function of tip length for a finite tungsten tip in a plane-wave excitation; the distance between the two peaks is close to the laser wavelength of $800\,\mathrm{nm}$, a clear sign of antenna resonance. The enhancement factor changes by about a factor of $2$ for different tip lengths. This shows that simulating a finite tip in a plane-wave excitation cannot give the correct field enhancement factor for a larger nanotip in a laser focus.}
    \label{antres}
\end{figure}

Our FDTD simulations of the field enhancement near nanotips were carried out using  Lumerical FDTD Solutions, a commercial Maxwell solver.
Our simulations encompass a cubic volume $V = X \times Y \times Z$ with the tip apex at the origin $\ort = 0$ and the tip shaft along the positive $x$ axis (Fig.~\ref{main-tipfield} is flipped $x \to -x$ with respect to the coordinates used in our simulation). The exact size of the volume depends on the parameters of a given simulation, as discussed below. As the volume that can be simulated with the FDTD algorithm is necessarily finite, care needs to be taken in the setup of the simulation to avoid unphysical antenna resonances due to the finite length of the simulated tip (see Fig.~\ref{antres} for an example)~\cite[S][]{Martin2001,Zhang2009}. We find that the results do not depend on the length of the tip and the size of the focus if one includes the focal spot inside the simulation volume and ensures that the laser's electric field at the simulation boundaries lateral to the propagation of the beam is negligible. We choose the size of the volume accordingly. Typical values are $X=Y=8000\,\nm$, $Z=1000\,\nm$.

The volume is meshed with a rectangular grid of non-constant resolution. At the tip apex, the resolution is considerably higher than in free space at a distance from the tip: the mesh node distance varies from approximately $50\,\mathrm{nm}$ in free space to $0.1\,\mathrm{nm}$ at the apex of the sharpest tip we simulate.

The laser is modeled as a Gaussian beam with the wave vector parallel to the $z$ axis and the polarization parallel to the $x$ axis and, thus, the tip shaft. The source area (i.e., the area where it enters the simulation) is at the negative $z$ boundary. In our time-domain simulation method,  we employ a  short laser pulse of duration $5\,\fs$ (intensity full width half maximum). Therefore, the laser light has a spectral width $\Delta\lambda$. We have verified in several tip geometries that the pulse duration has negligible effects on the field enhancement factor, so our results are also valid for longer pulses and continuous-wave excitations. This would be different for sharp resonances that critically depend on the wavelength. We did not observe such effects for the geometries under investigation.

The nanotip's optical properties are given by a dielectric function $\epsilon = \epsilon_\mathrm{r} + i \epsilon_\mathrm{i}$, which we obtain from experimental data samples of bulk metal~\cite[S][]{Haynes2011}. As with the pulse duration, the variance of $\epsilon(\lambda)$ for the spectral range of the laser pulse has no effect on field enhancement for the materials we studied in our simulations. This may be different for materials and wavelengths where $\epsilon(\lambda)$ varies rapidly, for example near bulk plasmon resonances. It should also be noted that a weak dependence of the dielectric constant on the structure size has been found for metal nanostructures smaller than the mean free path of conduction band electrons~\cite[S][]{Stoller2006}. As our simulations show, small changes of the dielectric constant do not significantly alter the field enhancement factor. Therefore, using the bulk dielectric constants in the simulation of nanotips should not result in significant systematic uncertainties.

A challenge for FDTD simulations of optical field enhancement are plasmonic tips (i.e., materials with $\epsr \ll -1$ and small $\epsi$, such as gold at $800\,\mathrm{nm}$) as they can cause a variety of numerical artifacts related to the appearance of surface plasmons~\cite[S][]{Novotny2006a, Raether1988}, which are excited at the apex and propagate along the tip shaft. Due to the rectangular FDTD mesh grid, the propagation of these plasmons is difficult to simulate (except for $\alpha = 0$) as they can scatter at the discrete steps of the material boundary, causing high loss. In some cases, such discretization errors can lead to unforeseen localized resonances along the tip shaft where electric fields may be `stuck' long after the laser pulse and surface plasmons are gone. Increasing the mesh resolution along the tip shaft does not prevent the appearance of such numerical artifacts due to the mismatch between the Cartesian grid and the local direction of the tip boundary. However, while these localized resonances hinder simulations of plasmon propagation on the conical shaft for $\alpha \neq 0$  which would be of importance for plasmonic nanofocussing~\cite[S][]{Stockman2004}, we found that the field enhancement factor at the apex could still be reliably calculated in almost all cases.
Only for sharp gold tips ($R < 10\,\nm$) with a small but non-zero opening angle near the plasmon resonance (Eq.~\ref{main-eq:sppres}), which exhibit the largest ``steps'' due to discretization errors discussed above, we observe an effect that prevents a correct simulation of near-field enhancement at the apex. At such tips, surface plasmons are coupled in at the shaft near the apex at the steps caused by discretization errors and propagate along the shaft from there, interfering with the near-field at the apex. This leads to an increased uncertainty for the field enhancement factor, and it can sensitively influence the phase shift.

The simulations were carried out on a desktop computer with an Intel Xeon CPU W3530 at $2.8\,\mathrm{GHz}$ and with $18\,\mathrm{GB}$ RAM. A single simulation typically took a few hours to complete. (This varied significantly depending on the simulation volume and the mesh resolution.) We exploit the symmetry of the setup with respect to reflection at the $y = 0$ plane to reduce computation time and memory requirements.

\section{FDTD: obtaining the field enhancement factor}

\begin{figure*}[bt]
    \begin{center}
    \includegraphics[width=\textwidth]{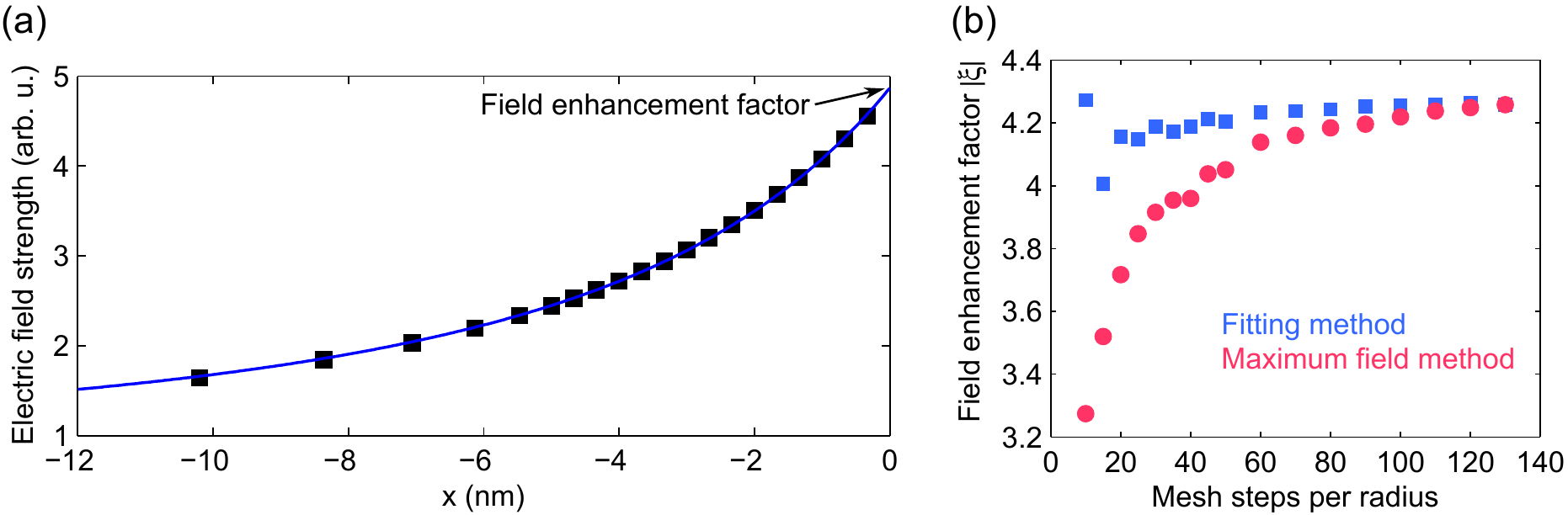}
    \end{center}
    \caption{(a) Example of the fitting method. The black dots are on-axis ($y = 0, z = 0$) simulation results of the electric field at the moment of the greatest enhancement. The blue line shows a fit using Eq.~\ref{fiteq}. (b) Field enhancement factor of a tungsten tip as as function of the mesh resolution near the apex, obtained by different methods: taking the maximum (red circles) and applying a quadratic fit (blue squares). Clearly, the fitting method is computationally less expensive. We typically use $30$ or $40$ mesh steps per radius for the simulations.}
    \label{xi}
\end{figure*}
The magnitude of the field enhancement factor $|\xi|$ is defined as the ratio of the maximally enhanced field strength to the driving field strength. The amplitude of the driving laser pulse in the bare focal plane ($z = 0$) is set to $1$ in our simulations. In principle, the field enhancement factor could therefore be obtained by simply taking the maximum of the electric field strengths $\efeld (\ort, t)$ in a simulation:
\begin{equation}
    |\tilde\xi| = \max_{x, y, z, t} | \efeld (\ort, t) |.
\end{equation}
However, there are several problems with this approach due to numerical limitations and artifacts. As the electric field strength decreases monotonically with distance from the tip surface, the maximum field strength is always found at a point of the simulation next to the material-vacuum boundary, and depends on the placement of the last grid point with respect to the boundary. Therefore, $|\tilde\xi|$ depends on the mesh resolution of the simulation at the boundary of the tip apex. A second problem arises due to stair-casing effects, which may cause an unrealistically high electric field strength at single points of the simulation. This effect is particularly noticeable for plasmonic materials.

To avoid the numerical problems related to simply taking the maximum, we use a more robust and efficient method to obtain $|\xi|$, as illustrated in Fig.~\ref{xi}. Note that the highest field enhancement occurs in the plane of symmetry if the laser polarization is parallel to the tip axis. Additionally, for the tips we investigate ($R \le 30\,\nm$ and $\lambda = 800\,\nm$), the maximum is at or very close to the tip axis $y=z=0$. It is therefore sufficient to analyze the on-axis electric fields $\efeld (x, 0, 0, t)$ in order to obtain the field enhancement factor. The deviation in strength from the actual field maximum is around $6\,\%$ for $30\,\nm$ radius and less than $1\,\%$ for $5\,\nm$ radius. If we investigated larger tip radii or, equivalently, smaller wavelengths, the maximum field strength would shift further away from the axis~\cite[S][]{Yanagisawa2010} and we would have to take this asymmetry into account.

We obtain the field enhancement factor in the following way. First, we find out the time of the greatest enhancement $t_\mathrm{max}$ by locating $\max_t | \efeld(\ort', t) |$ at a point $\ort'$ close to the tip apex. Then, we consider the electric field at $t=t_\mathrm{max}$ on the $y=z=0$ line outside the tip ($x<0$) and fit a quadratic decay
\begin{equation}
    f(x) = \frac{a}{(x - x_0)^2} + f_\mathrm{bg}
    \label{fiteq}
\end{equation}
to it. We extrapolate the fit function back to the tip surface at $x=0$, and the value of the fit function at this point yields $|\xi|$. In the fit function, the $1/(x-x_0)^2$ term models the near-field and $f_\mathrm{bg} = \cos(\phi)$ is the background field strength of the exciting laser pulse. While the background field amplitude is $1$, $f_\mathrm{bg}$ also takes the phase shift $\phi$ between near-field and exciting field into account. For the phase shift, see below. $a$ and $x_0$ are the free fit parameters. An example of such a fit is shown in Fig.~\ref{xi}(a). Note that we only evaluate the fit function and the simulation results on a line that is much smaller than the waist radius $w_0$, so we can assume the background field strength to be constant.

It should be noted that it is not clear from the simulations that the near-field decreases quadratically with distance. In fact, fit functions with powers of $1$ to $3$ produce an almost equally good fit and yield approximately the same field enhancement factor. If the power itself is allowed to vary in the fit, we obtain non-integer powers between $1$ and $3$, with different results for different simulations. This is unlike the near-field at nanospheres, for example, which shows a third-order decrease with the singularity exactly at the center of the sphere~\cite[S][]{Maier2007}. We have chosen a quadratic fit function because it leads to a position of the singularity $x_0$ close to the center of the sphere at the tip apex. In any case, the choice of fit function changes $\xi$ only insignificantly (by ${\sim}1.5\,\%$ in the example of Fig.~\ref{xi}(a)).

A comparison of the enhancement factors obtained by fitting and by simply taking the maximum is shown as a function of the mesh resolution in Fig.~\ref{xi}(b). The field maximum converges much more slowly than the quadratic fit method, which deviates by less than $5\,\%$ from the final value of $\xi$ even for low resolutions, i.e., few mesh steps per radius. They both converge to the same value. This shows that stair-casing effects do not cause unrealistically high field strengths in this series of simulations. 

We conclude that the near-field around the tip apex is already modeled correctly at lower resolutions (${\sim}40$ steps per radius) and that the additional dependence on the mesh resolution comes only from the discretization of the mesh, which we can efficiently circumvent by extrapolating the near-field to the surface of the tip as described above. As a compromise between precision of the results and computational resource requirements, we used a mesh resolution of $R / 40$ as function of tip radius $R$ for all simulations except the ones where we vary the dielectric function of the tip (Fig.~\ref{main-epsxi}). There we used a mesh resolution of $R / 30$ to speed up the computations.

The phase shift $\phi$ can be obtained by comparing the zero-crossing of the near-field close to the tip with the zero-crossing of the undisturbed pulse at negative $x$. Due to the limited temporal resolution of our simulations and numerical dispersion~\cite[S][]{Taflove2005} that shifts the carrier-envelope phase in a mesh-dependent way, this method comes with an unavoidable error, which we estimate to be around $\Delta\phi \approx 0.05\pi$ by comparing simulations of the same nanotip with different mesh resolutions and simulation volumes. With knowledge of both the phase shift and the magnitude of the enhancement, we can completely characterize $\xi = |\xi| \exp(i\phi)$.

In a final step, we apply a correction $\xi \to \xi / 0.95$ to the field enhancement factor. The value of $0.95$ is obtained from simulations of the laser pulses without including the nanotip. This correction factor compensates pulse propagation effects in the simulation, which reduce the amplitude of the exciting pulse in the focal plane. We attribute these effects to both numerical dispersion and our use of Gaussian pulses in a regime where the waist radius $w_0$ is of the same order of magnitude as the wavelength  $\lambda$. 

\section{Boundary element method}

To rule out systematical errors from the space discretization and time integration in the FDTD simulations, we double-checked the reliability of our simulations by also numerically solving Maxwell's equations with the boundary element method (BEM) as implemented in the public-domain SCUFF-EM package~\cite[S][]{SCUFFEM, Reid2013a, Reid2013}. Being a frequency-domain method, the boundary element method is free from time integration errors that contribute to the errors in FDTD. Time-domain quantities can be reconstructed by superimposing many frequency components and the convergence of this Fourier synthesis can be checked by increasing the frequency range and resolution. The boundary element method takes advantage of the analytically known solutions of Maxwell's equations in homogeneous media, so that only the surface of the tip is discretized. This can lead to lower memory requirements and improved scaling compared to FDTD, where the three-dimensional simulation volume must be discretized. Importantly, this smooth discretization of the tip surface also allows us to assess the influence of the Cartesian grid that is employed in the FDTD simulations leading to staircasing artifacts.

\begin{figure*}[t]
  \includegraphics[width=\textwidth]{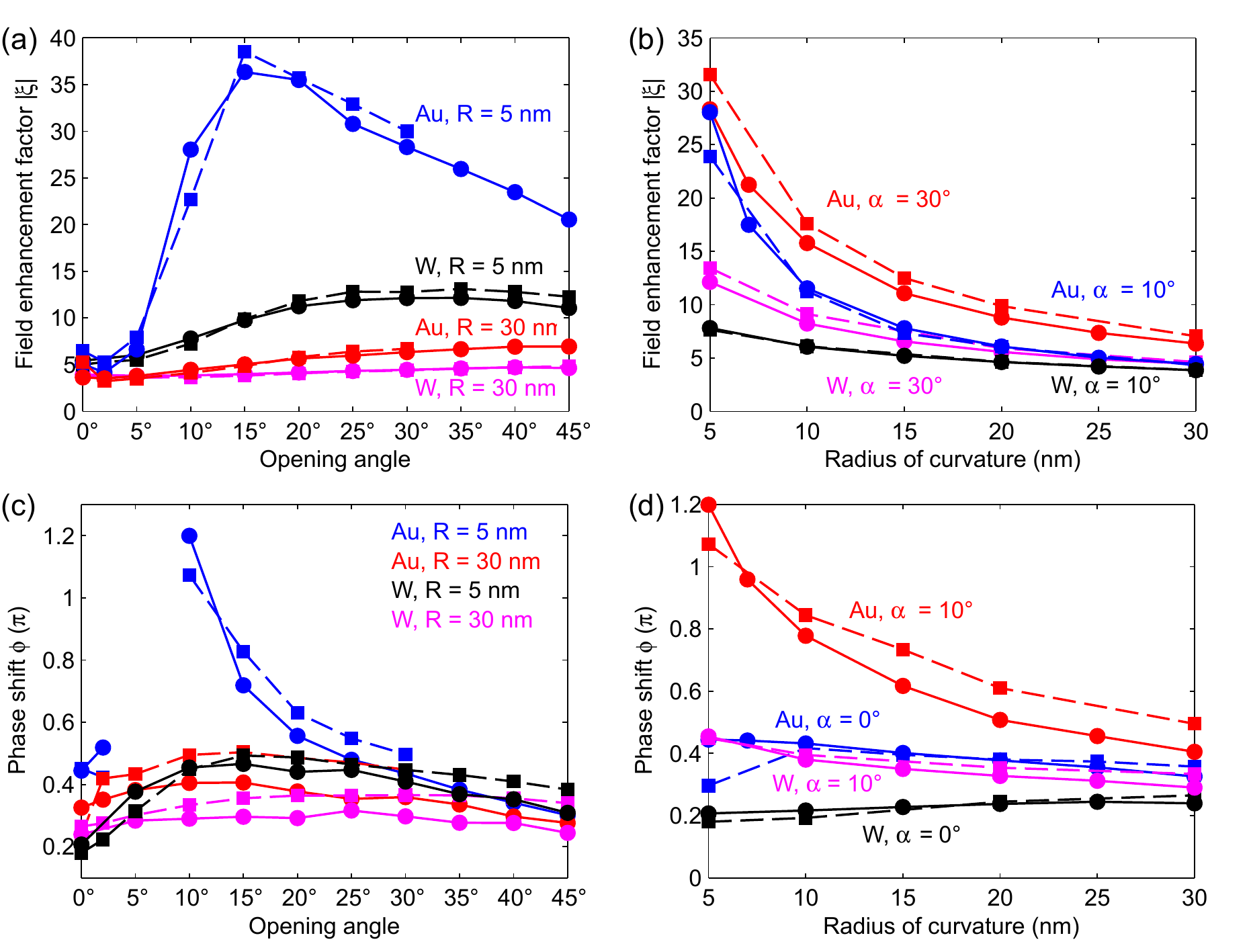}
  \caption{Comparison between FDTD results (circles connected by solid lines) and BEM results (squares connected by dashed lines) for the field enhancement factor $|\xi|$ (a,b) and phase shift $\phi$ (c,d) in different geometries. The missing values for the phase shift of $R=5\,\nm$ gold tips around $\alpha=5\degree$ are due to the numerical problems with this geometry, as discussed in appendix~\ref{fdtdsetup}.}
  \label{bemfdtd}
\end{figure*}
 
\begin{figure*}[p]
    \includegraphics[width=\textwidth]{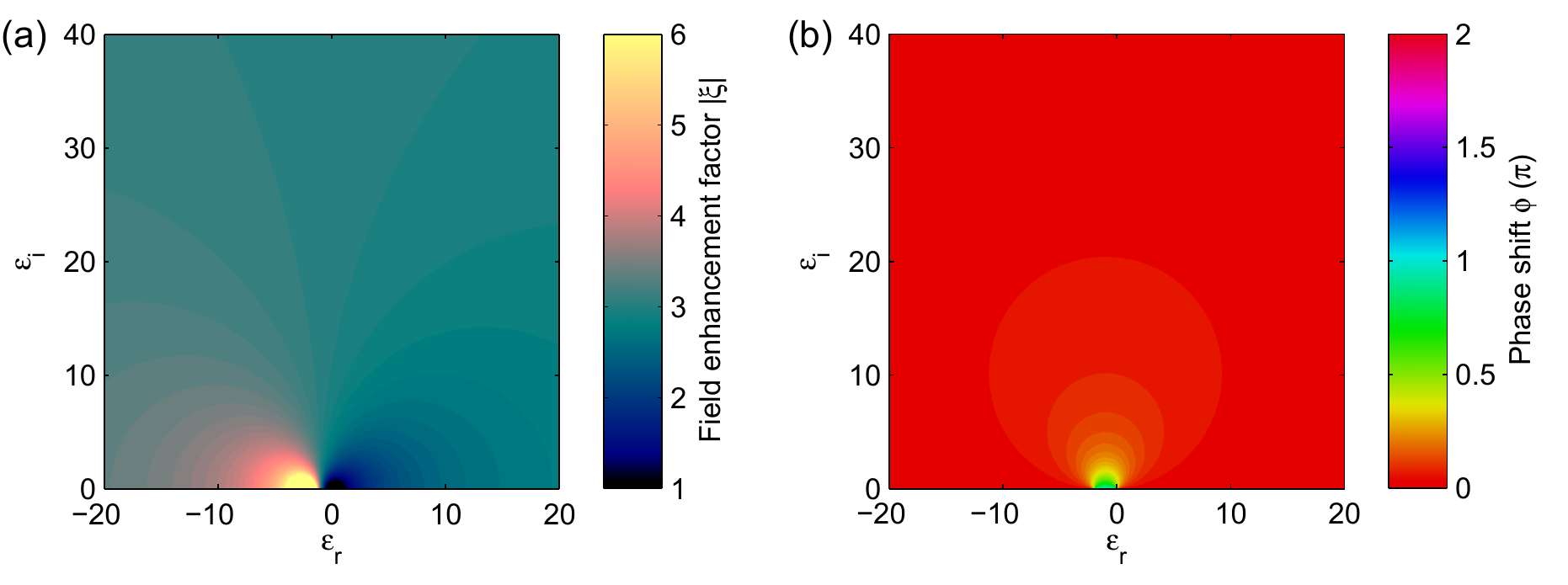}
    \caption{Complex field enhancement factor $\xi = |\xi| \exp(i \phi)$ of nanospheres (aspect ratio $r = 1$, shape factor $A = 1/3$) with radius $R \ll \lambda$ obtained from Eq.~\ref{ellipsexi}. (a) $|\xi|$, (b) $\phi$. }
    \label{nanospheres}
\end{figure*}

\begin{figure*}[p]
    \includegraphics[width=\textwidth]{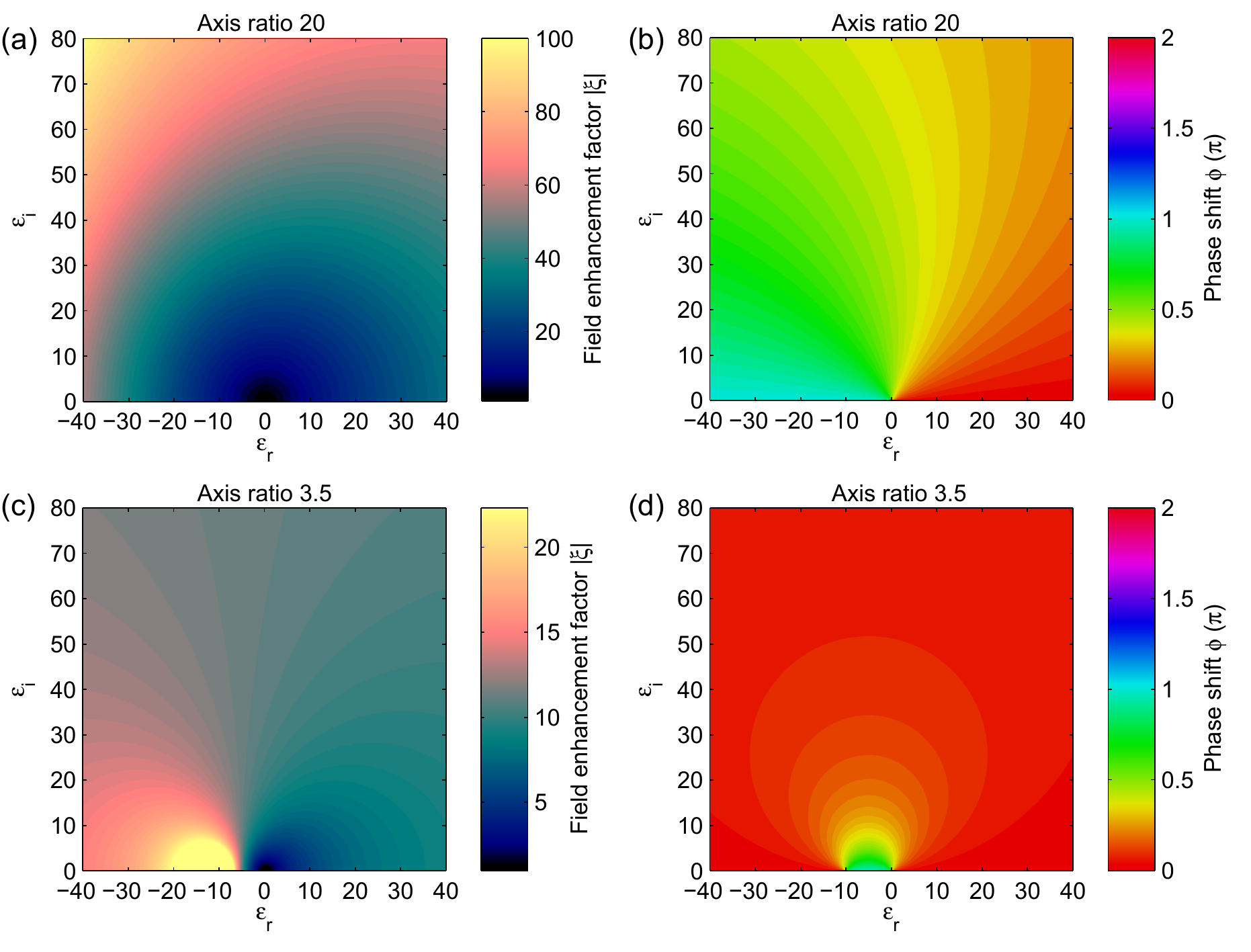}
    \caption{Complex field enhancement factor $\xi = |\xi| \exp(i \phi)$ of ellipsoids with aspect ratios $20$ (a,b) and $3.5$ (c,d) as a function of the dielectric constant $\epsilon = \epsr + i \epsi$, according to Eq.~\ref{ellipsexi}. Left column: $|\xi|$, right column: $\phi$. }
    \label{xiellipse}
\end{figure*}

A typical simulation run proceeds as follows. The tip geometry is defined depending on the geometrical parameters tip radius and opening angle as for the FDTD calculations. First, the surface of the tip is discretized into $\npa$ triangles employing the public-domain meshing software gmsh~\cite[S][]{Geuzaine2009}. We use an adaptive mesh to resolve the small-scale features of the near-field around the tip apex with discretization steps of $0.2\,\nm$ near the apex. The remainder of the tip is discretized in larger steps of about $1\, \nm$ to $20\,\nm$ that resolve the geometry of the tip and are much smaller than the wavelength of surface plasmons that can be excited at the sharp tip apex. The total length of the simulated tip was between $1.7$ micron and $6$ micron. The inside of the tip is designated the experimental dielectric bulk constant of the material at the working wavelength~\cite[S][]{Palik1991}. The incident field is chosen as a focused laser beam as for the FDTD results~\cite[S][]{Sheppard1999}. The boundary element method solver SCUFF-EM is then employed to solve for the electromagnetic fields where the numerical cost scales with the size of the BEM matrix, ${\sim}\npa^2$. For the calculations presented in this paper, $\npa \approx 10000$ which corresponds to ${\sim}15\,\mathrm{GB}$ of RAM. After the BEM matrix equations are solved by standard linear algebra methods, the electric near-field in the region $0.05\,\nm$ in front of the tip axis is evaluated and extrapolated to the tip apex. The field enhancement and phase shift are then given by the absolute value and phase of the ratio of the total field perpendicular to the tip surface to the incoming field along the tip axis. The field enhancement is only weakly dependent on the laser wavelength so that the phase shift corresponds to a carrier-envelope phase shift for few-cycle laser pulses when the time-dependent near field is reconstructed by a Fourier transform.

The boundary element method is restricted to piecewise homogeneous material configurations, so that absorbing boundaries like perfectly matched layers that exist for FDTD or finite element methods are precluded. This can lead to problems for materials where the propagation length of surface plasmons on the structure of interest is larger than the size of the structure that can be modeled. For tungsten, which has a large imaginary part of the dielectric function around $800\,\nm$, excitations from the tip apex propagating  along the tip shaft decay rapidly (typically within $200\,\nm$~\cite[S][]{Sarid2010}). However, the situation changes for plasmonic materials like gold 
where the propagation distance of surface plasmons can be up to several tens of microns, rendering the simulation of the mesoscopic structure up to the length where the plasmons are fully decayed numerically infeasible. We instead use tips of a few micron length also for plasmonic materials and exploit the fact that, for short enough pulses, the incident and reflected electric fields are well separated in time. In frequency space, the reflections of surface plasmons from the back end of the tip contribute to the near-field at the tip apex, leading to unphysical peaks in the electric near-field at frequencies that change for different tip lengths (``antenna resonances''). We filter out the contributions of the reflected surface plasmons by transforming to the time domain and only taking into account the short-time response to a few-cycle laser pulse, as the surface plasmon wave packet that is reflected from the back end of the tip will be delayed by at least $7 \,\fs$ per micron tip length (speed of light $c \approx 300\,\nm/\fs$). We find that, while the interference pattern stemming from the antenna resonances changes with increasing tip length, the short-time behavior calculated by a Fourier transform of the laser spectrum is well converged if the incident and reflected wave packets are well separated in time, which can be achieved by a tip length substantially below the surface plasmon propagation length. This low-pass filter in the time domain corresponds to filtering out the high-frequency oscillations of the antenna resonances in frequency space, i.e., smearing out the interference fringes over the spectrum of a short incident laser pulse.
The BEM calculations for plasmonic materials, where simulations at several wavelengths must be combined, are thus significantly more costly than those for non-plasmonic materials.

\section{Comparison between FDTD and BEM results}
\label{comparison}

In Fig.~\ref{bemfdtd}, we compare results for the field enhancement factor and phase shift of nanotips obtained from simulations using either the finite-difference time-domain method (FDTD) or the boundary element method (BEM). Shown here are results for different geometries of tungsten and gold tips. In general, we find a good agreement between the two numerical methods. As discussed above, gold tips are more challenging to simulate than tungsten tips for both the FDTD and BEM methods, so it is not surprising that the agreement between the two methods is somewhat better for tungsten than for gold.

The field enhancement factor obtained by the two methods typically agrees within ${\sim}10\,\%$, with the exception of a few particular geometries in the vicinity of the  plasmon resonance like $(R=5\,\nm, \alpha = 10\degree, \mathrm{Au})$ in Fig.~\ref{bemfdtd}(a), where we observe deviations of around $20\,\%$. For the phase shift, the deviation between the two methods is approximately $0.1\,\pi$.

We conclude that the results presented in this article do not exhibit significant systematic errors due to the choice of simulation method, and that both FDTD and BEM are well suited for the simulation of near-fields at nanotips.

\section{Comparison to nano-ellipsoids} 

\begin{figure}[t]
  \begin{center}
    \includegraphics[width=\columnwidth]{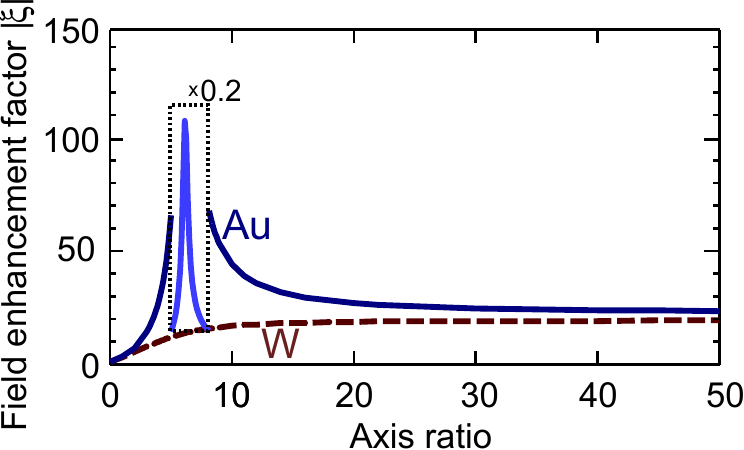}
  \end{center}
    \caption{Field enhancement factor $|\xi|$ of tungsten (red dashed) and gold (blue solid) nano-ellipsoids at $\lambda=800\,\nm$ (Eq.~\ref{ellipsexi}). For better visibility, the field enhancement of gold between $5 \degree$ and $8 \degree$ is scaled by $0.2$ (dotted box). 
    }
    \label{elliauw}
\end{figure}

To elucidate the relationship between field enhancement and dielectric function, we compare our simulations for nanotips to the near-field of ellipsoids 
for which an analytic solution is available in the static limit~\cite[S][]{Sarid2010, Martin2001, Bohren2008}, see also \cite[S][]{Novotny2006a, Neacsu2005a}. For a rotationally symmetric ellipsoid with two equal axes $b=c$ and a major axis $a$ along the polarization direction, the complex field enhancement factor for a given $\epsilon(\lambda)$ is (in the limit $a, b, c \ll \lambda$) 
\begin{equation}
 \xi(\lambda) = \frac{ \epsilon(\lambda) }{1 + \left[ \epsilon(\lambda) - 1 \right] A(r)}
 \label{ellipsexi}
\end{equation}
with the so-called shape factor $A(r)$ depending on its aspect ratio $r = a/b$, 
\begin{equation}
 A(r) = 
 \frac{1}{1-r^2} - \frac{r \,\, \arcsin\left(\sqrt{1-r^2}\right)}{(1-r^2)^{3/2}}  .
 \label{ellipseA}
\end{equation}
The shape factor varies smoothly from $A(r \to 0) =1$ for pancake-like oblate ellipsoids via $A(r=1) = 1/3$ for spheres to $A(r \to \infty) =0$ for cigar-like prolate ellipsoids. 
The resulting field enhancement (Eq.~\ref{ellipsexi}) assumes its minimum around $\epsilon \to 0$ while its maximum is found at the dipole resonance at the pole of Eq.~\ref{ellipsexi}, i.e., for
\begin{equation}
\label{eq:feell}
\epsilon = 1 - 1/A(r) \,\, .
\end{equation}
Eq.~\ref{eq:feell} encodes the relationship between dielectric function and geometry in analogy to Eq.~\ref{main-eq:sppreseps} with the aspect ratio playing a similar role as the tip opening angle in Eq.~\ref{main-eq:sppreseps}. For nano-spheres ($A(r=1) = 1/3$, Fig.~\ref{nanospheres}), we find $\epsilon = -2$, thereby recovering the first Mie plasmon at $\omega = \omega_p / \sqrt{3}$ for a Drude metal ($\epsilon_\mathrm{Drude}(\omega) = 1 - \omega_p^2 / \omega^2$). Away from the resonance, the field enhancement for a nanosphere asymptotically approaches $\xi(|\epsilon| \to \infty) = 1/A(1) = 3$. 

For other aspect ratios, the overall shape of $\xi(\epsilon)$ remains the same while its value $\xi(|\epsilon| \to \infty)$ changes. For any aspect ratio, a resonance is only attainable for materials with a negative dielectric function $\mathrm{Re}(\epsilon) < 0$. The transition from a sphere to a needle-like ellipsoid changing the shape factor from $A=1/3$ to $A \to 0$ in Eq.~\ref{ellipsexi} magnifies the region of appreciable field enhancement. This is illustrated by comparing Fig.~\ref{nanospheres} with Fig.~\ref{xiellipse}c,d, which shows $\xi(\epsilon)$ for an elongated ellipsoid with aspect ratio 3.5. As the aspect ratio increases, the position of the resonance moves to more negative values of $\mathrm{Re}(\epsilon)$. 

The field enhancement factor of a needle-like ellipsoid with a large aspect ratio $r = 20$ (Fig.~\ref{xiellipse}(a,b)) resembles the extreme case $r \to \infty$, where the field enhancement factor is simply $\xi(\epsilon) = \epsilon$. The same result was found for paraboloids in the quasi-static approximation~\cite[S][]{Chang2009}. The increasing enhancement of the electric field with increasing discontinuity of $|\epsilon|$ at the ellipsoid's boundary can be interpreted as broadband field enhancement due to the lightning rod effect~\cite[S][]{Thomas2013}. The other extreme case of a pancake-like surface, $r = 0$, yields a vanishing field enhancement $\xi(\epsilon) = 1$.

The near-field at nano-ellipsoids is qualitatively similar to nanotips, with the aspect ratio of the ellipsoid playing a role analogous to the opening angle of the tip. Comparing Fig.~\ref{main-epsxi} and Fig.~\ref{xiellipse}, we find that slim nanotips $\alpha = 0\degree$ behave similarly to slim ellipsoids with aspect ratio $20$ (increasing enhancement factor with $|\epsilon|$, increasing phase shift for larger angles $\mathrm{arg}(\epsilon)$), while broader nanotips with opening angle $\alpha = 30\degree$ are similar to broader ellipsoids with aspect ratio ${\sim}3.5$ (broad plasmon resonance in the $\epsr < 0$ region, large phase shift in between $0$ and the resonance).

The angle dependence of tungsten and gold tips (Fig.~\ref{main-fig:angle}a) may be compared to the aspect ratio dependence of tungsten and gold ellipsoids (Fig.~\ref{elliauw}). The latter show low field enhancement for small aspect ratios and converge to approximately the same enhancement factor of ${\sim}20$ for high aspect ratios as they share a similar value of $|\epsilon|$ at $\lambda=800\,\nm$ (see Fig.~\ref{main-fig:dielfct}). In between, however, the behavior is different: While the field enhancement factor of tungsten increases monotonically, gold exhibits an additional plasmon resonance at an aspect ratio of around $r=6$, leading to far higher field enhancement.

\providecommand{\newblock}{}